\documentclass[12pt]{article}
\usepackage{palatino,cite,epsfig}
\usepackage{graphics,graphicx}

\newlength{\capindent}
\newlength{\capwidth}
\newlength{\figwidth}
\newcommand{\icaption}[2][!*!,!]{\hspace*{\capindent}%
  \begin{minipage}{\capwidth}
    \ifthenelse{\equal{#1}{!*!,!}}%
      {\caption{#2}}%
      {\caption[#1]{#2}}
  \end{minipage}}
\def\a34{\cos\alpha_{34}}

\def\ee{\mathrm{e^{+}e^{-}}}
\def\AFB{\mathrm{\rm A_{FB}\;}}

\def\TO{\mathrm{\longrightarrow\ }}

\def\GG{\mathrm{\rm \;\gamma\gamma\;}}

\def\L{\mathrm{\Lambda}}
\def\SP{\mathrm{s^{\prime}}}

\def\rs{\sqrt{s}}

\def\xi{x_{i}}

\newcommand{\GeV}{\ensuremath{\mathrm{Ge\kern -0.12em V}}}
\newcommand{\TeV}{\ensuremath{\mathrm{Te\kern -0.12em V}}}
\def\LL{\mathrm{LL}}
\def\RR{\mathrm{RR}}
\def\RL{\mathrm{RL}}
\def\LR{\mathrm{LR}}

\newcommand {\Be}{\begin{equation}}
\newcommand {\Ee}{\end{equation}}

\newcommand {\Eqref}[1]{Equation~(\ref{#1})}

\newcommand {\Figref}[1]{Figure~\ref{fig:#1}}

\newcommand {\Tabref}[1]{Table~\ref{tab:#1}}

\begin{document}

\vspace{1cm}

\begin{flushright} 
hep-ph/0305125 \\
May 2003
\end{flushright}
\vspace{0.5cm}

\begin{center}

{\Large\bf Sensitivity to Contact Interactions and Extra Dimensions in
Di-lepton and Di-photon Channels at Future Colliders}

\vspace{0.4cm}

{\bf Contribution to LHC / LC Study Group Working Document%
\footnote{
For further informations, see {\tt
http://www.ippp.dur.ac.uk/$\sim$georg/lhclc/ }. For questions and comments,
please contact {\tt Georg.Weiglein@durham.ac.uk }.}
}
 
\vspace{1cm}

{\sc 
Dimitri Bourilkov
}

{\small bourilkov@mailaps.org
}

\vspace*{1cm}

{\sl
    University of Florida\\
    Gainesville, FL, USA

\vspace*{0.4cm}
}
\end{center}

\vspace*{1cm}

\begin{abstract}
Virtual effects from a generic description of physics beyond the Standard
Model in terms of contact interactions, or from large extra dimensions
will modify the observed cross sections for easy to detect final states
like lepton or photon pairs, and can be used to probe scales much higher 
than the center-of-mass energy of the partons initiating the interactions. 
In this note the sensitivity reach of the Large Hadron Collider to contact
interactions in the Drell-Yan channels and of a Future Linear Collider 
to contact interactions and extra dimensions in $e^+e^-$, $\mu^+\mu^-$
and $\gamma\gamma$
final states are studied. Experimental aspects of the measurements,
systematic error effects and the evolution of the search reach with
accumulated luminosity are considered.

\end{abstract}

{\it Work presented at the LHC/LC Study Group (CERN 5 July 2002, 14 February and 9 May 2003)}



\section{Introduction}

The Large Hadron Collider (LHC) and a Future Linear Collider (FLC)
open complementary possibilities at the high energy frontier. The latter offers
the benefits of a well defined
initial state while the former can be viewed as a wide band parton beam,
capable of probing deep into the $\TeV$ region. Many new effects, if discovered
at LHC, can be studied in greater detail at a FLC.

The most desirable case is the direct observation of unknown physics
phenomena, e.g. a peak in a mass spectrum. Even if we are not so lucky,
virtual effects from a generic description of physics beyond the Standard
Model (SM) in terms of contact interactions, or from large extra dimensions
will modify the observed cross sections for easy to detect final states
like lepton or photon pairs, and can be used to probe scales much higher 
than the center-of-mass energy $\rs$ of the partons initiating the interactions. 

There have been many studies of the sensitivity reach of LHC
and a FLC to contact interactions or extra dimensions.
Here we will concentrate on some experimental aspects of the measurements, the
effects of systematic errors and the evolution of the search reach with
accumulated luminosity.

\section{Contact Interactions}

Contact interactions offer a general framework for describing 
a new interaction with typical energy scale  $\L \gg \rs$.
The presence of operators with canonical dimension
$N > 4$ in the Lagrangian gives rise to effects  $\sim 1/M^{N-4}$.
Such interactions can occur for instance, if the SM particles are composite, or
when new heavy particles are exchanged.

\begin{table}[htb]
\renewcommand{\arraystretch}{1.20}
\caption{Contact interaction models.}
\label{tab:ci-models}
  \begin{center}
    \begin{tabular}{|c|cccc|cccc|}
\hline
~~Model~~&~~~LL~~&~~~RR~~&~~~LR~~&~~~RL~~&~~~VV~~
&~~~AA~~&~LL+RR~&~LR+RL~\\
        & \multicolumn{4}{c|}{ Non-parity conserving } & \multicolumn{4}{c|}{ Parity conserving } \\
  \hline \hline
$\eta_{\LL}$
   & $\pm$1& 0    &    0 &    0 &$\pm$1&$\pm$1&$\pm$1&  0   \\
$\eta_{\RR}$
   & 0     &$\pm$1&    0 &    0 &$\pm$1&$\pm$1&$\pm$1&  0    \\
$\eta_{\LR}$
   & 0     & 0    &$\pm$1&    0 &$\pm$1&$\mp$1& 0   &$\pm$1  \\
$\eta_{\RL}$
   & 0     & 0    &    0 &$\pm$1&$\pm$1&$\mp$1& 0   &$\pm$1 \\
\hline
    \end{tabular}
  \end{center}
\end{table}

In the following we will consider fermion- or photon-pair production.
In the fermion case, the lowest order flavor-diagonal and helicity-conserving
operators have dimension six~\cite{PeskinCI}.
For photons the lowest order operators have dimension eight, leading to
suppression of the interference terms as the inverse fourth power of the
relevant energy scale.

The differential cross section takes the form
\Be
\frac{d \sigma}{d \Omega} = SM(s,t)+\varepsilon\cdot C_{Int}(s,t)+\varepsilon^2\cdot C_{NewPh}(s,t)
\Ee
where the first term is the Standard Model contribution, the second comes from
interference between the SM and the contact interaction, and the third is the
pure contact interaction effect.
The Mandelstam variables are denoted as $s$, $t$ and $u$.

Usually the coupling is fixed,
and the structure of the interaction is parametrized by coefficients
for the helicity amplitudes:
\begin{center}
\begin{tabular}{ll}
 g            &  coupling (by convention $\frac{g^2}{4\pi}=1$) \\
 $|\eta_{ij}|\leq 1$  & helicity amplitudes ($i,j = \rm L,R$)\\
 $\varepsilon$  & $\frac{g^2}{4\pi}\frac{sign(\eta)}{\L^2}$ for $f\bar f$; $\ \ \sim \frac{1}{\L^4}$ for $\GG$ 
\end{tabular}
\end{center}

Some often investigated models are summarized in~\Tabref{ci-models}. The models
in the second half of the table are parity conserving, and hence not
constrained by the very precise measurements of atomic parity violation at low
energies.
The results presented in this contribution cover the models in the table.

The discovery reach for a given model is determined by a fitting procedure
similar to the one used for the analysis of LEP2 data on Bhabha
scattering~\cite{Bourilkov:1999iz,Bourilkov:2000ap}.
A negative log-likelihood function is constructed by combining
all simulated data points:
\begin{equation}
-\log {\cal{L}} = \rm \sum_{r=1}^{n}\left(\frac{[prediction(SM, \varepsilon) - measurement]^2}{2 \cdot \Delta^2}\right)_r
\label{eqll1}
\end{equation}
where $\rm prediction(SM, \varepsilon)$ is the SM expectation for a
given measurement (cross section or forward-backward asymmetry)
combined with the additional effects of new physics as a function of
their characteristic scale,
$\rm measurement$ is the corresponding measured (here simulated) quantity and
\mbox{$\Delta = \rm error[prediction(SM, \varepsilon) - measurement]$}.
The index $r$ runs over all data points.
For contact interactions
\begin{equation}
\varepsilon = \frac{1}{\Lambda^2}.
\label{eqepsci}
\end{equation}
The error on a deviation consists of three parts, which are combined in
quadrature: a statistical error, a systematic error (our best guess) and 
a theoretical uncertainty (expected to be in general quite small, but
still important for large accumulated luminosities).
One-sided lower limits (i.e. the sensitivity) on the scale
(e.g. $\Lambda$ for contact interactions) at 95\% confidence level are
derived for the two possible signs of the interference terms.
This is done by integrating the log-likelihood functions
in the physically allowed range of the parameters describing
new physics phenomena, assuming a uniform prior distribution.

\subsection{Drell-Yan at the Large Hadron Collider}

In the Standard Model the production of lepton pairs in
hadron-hadron collisions, the Drell-Yan process~\cite{DrellYan},
is described by s-channel exchange of photons or Z bosons.
The parton cross section in the center-of-mass system has the form:
\begin{equation}
\rm \frac{d \sigma}{d \Omega} = \frac{\alpha^2}{4s} [A_0(1+\cos^2\theta)+A_1\cos\theta]
\label{llnewph}
\end{equation}
where $\rm  \sigma = \frac{4 \pi \alpha^2}{3s} A_0$, $\rm \AFB = \frac{3}{8}\frac{A_1}{A_0}$ give the total cross section and the forward-backward
asymmetry. The terms $\rm A_0$ and $\rm A_1$ are fully determined by
the electroweak couplings of the initial- and final-state fermions.
At the Z peak the Z exchange is dominating and the interference term
is vanishing. At higher energies both photon and Z exchange contribute
and the large value of the forward-backward asymmetry is due to
the interference between the neutral currents.

Fermion-pair production above the Z pole  is a
rich search field for new phenomena at present and future high
energy colliders.
The differential cross section is given by
\begin{equation}
\rm \frac{d \sigma}{d \Omega} = |\gamma_s+Z_s +  New\;Physics\;?!|^2
\label{llxsec}
\end{equation}
where many proposed types of new physics can lead to observable
effects by adding new amplitudes or through their interference
with the neutral currents of the SM.

At hadron colliders the parton cross sections are folded with the
parton density functions (PDF): \mbox{$\rm pp \TO l_1l_2$}
\begin{equation}
\rm \frac{\mbox{d}^2\sigma}{\mbox{d}M_{ll}\mbox{d}y} 
[pp\rightarrow l_1 l_2] \sim \sum_{ij} 
\, \left(f_{i/p}(x_1)  f_{j/p}(x_2) +(i \leftrightarrow j)\right)\,  \hat{\sigma}  
\label{llpdf1}
\end{equation}
$\rm \hat{\sigma}$ - cross section for the partonic subprocess $ij\rightarrow l_1 l_2$\\
$\rm M_{ll}=\sqrt{\tau s} = \sqrt{\hat{s}}$ - mass, $\rm y$ - rapidity of the lepton pair\\
$\rm x_1=\sqrt{\tau}e^y$, $\rm x_2=\sqrt{\tau}e^{-y}$ - parton momentum fractions,\\
$\rm f_{i/p(\bar{p})}(x_i)$ - probability to find a parton $i$ with momentum
fraction $x_i$ in the (anti)proton.
\begin{eqnarray}
\rm \sigma_{F\pm B}(y,M) & = & [\int_{0}^{1} \pm \int_{-1}^{0}]\sigma_{ll} d(\cos \theta^{*}) \\
\rm \AFB (y,M) & = & \frac{\sigma_{F-B}(y,M)}{\sigma_{F+B}(y,M)}.
\label{llpdf2}
\end{eqnarray}

\begin{table}[htb]
  \begin{center}
\caption{x$_1$ and x$_2$ for different masses and rapidities.}
\label{tab:xmasrap}
\vskip0.2cm
\begin{tabular}{|c|c|c|c|}
\hline
   y      &     0         &     2       &      4        \\
\hline
             \multicolumn{4}{|c|}{M = 91.2 GeV}          \\
\hline
$\rm x_1$ &   0.0065      &   0.0481    &    0.3557     \\
$\rm x_2$ &   0.0065      &   0.0009    &    0.0001     \\
\hline
             \multicolumn{4}{|c|}{M = 200  GeV}          \\
\hline
$\rm x_1$ &   0.0143      &   0.1056    &    0.7800     \\
$\rm x_2$ &   0.0143      &   0.0019    &    0.0003     \\
\hline
             \multicolumn{4}{|c|}{M = 1000 GeV}          \\
\hline
$\rm x_1$ &   0.0714      &   0.5278    &       -       \\
$\rm x_2$ &   0.0714      &   0.0097    &       -       \\
\hline
    \end{tabular}
  \end{center}
\end{table}

The total cross section and the forward-backward asymmetry are
function of observables which are well measured experimentally
for $e^+e^-$ and $\mu^+\mu^-$:
the invariant mass and the rapidity of the final state lepton-pair.
This allows to reconstruct the center-of-mass energy of the
initial partons, even if their flavors are unknown. This will
be used in the subsequent analysis of contact interactions
by performing a scan of the high mass region above 0.5~$\TeV$.
For a ($\rm x_1 \geq x_2$) pair of partons  we have 4 combinations of
{\em up-} or {\em down-}type
quarks initiating the interaction:
$u\bar u, \bar u u, d\bar d, \bar d d$.
In $pp$ collisions the antiquarks come always from the sea and the quarks
can have valence or sea origin. The x-range probed depends on the mass
and rapidity of the lepton-pair as shown in Table~\ref{tab:xmasrap}.
Going to higher rapidities increases the difference between
x$_1$ and x$_2$ and hence the probability that the first quark is
a valence one. This allows a measurement of the forward-backward asymmetry
even for the symmetric initial $pp$ state.

Events are generated with the {\tt PYTHIA~6.2}$\;$
Monte Carlo~\cite{pitia} (with default PDF CTEQ5L) 
by applying the following cuts for both leptons:
\begin{itemize}
 \item pseudorapidity $\rm |\eta| < 2.5$
 \item transverse momentum $\rm p_T > 20$ GeV
\end{itemize}
which cover the barrel and endcaps of a typical LHC detector. The backgrounds
for these final states are low and can be suppressed further by isolation cuts.
It is interesting
to enlarge the acceptance to the very forward region. Experimentally this
is very demanding and will not be considered here. Everywhere in this study
only acceptance cuts are applied and the experimental efficiency within this
region is kept at 100 \%. For lower efficiency the luminosity for obtaining the
same result has to be rescaled accordingly. The results presented here
are an extension of the studies for the LHC SM workshop
(see~\cite{Haywood:1999qg,dbcmsnote:2000} and references therein), using much
higher statistics and applying a rigorous statistical procedure to determine the
sensitivity reach.

\begin{table}[hbt]
\renewcommand{\arraystretch}{1.20}
\caption{Sensitivity reach in one experiment for contact interactions
(LL model) at 95\% CL at LHC.}
\label{tab:dyci-limits}
 \begin{center}
  \begin{tabular}{|c|c|c||c|c|}
   \hline
   \multicolumn{5}{|c|}{\boldmath $pp \rightarrow e^{+}e^{-}X,\ \mu^{+}\mu^{-}X\ \ $ One Experiment \unboldmath} \\
   \multicolumn{5}{|c|}{\boldmath Contact Interactions LL Model \unboldmath} \\
 & \multicolumn{2}{c}{\boldmath 3 \% syst. err. \unboldmath} & \multicolumn{2}{c|}{\boldmath 6 \% syst. err. \unboldmath} \\
   \hline
Luminosity &  $\Lambda^{-}$  & $\Lambda^{+}$ &  $\Lambda^{-}$ & $\Lambda^{+}$ \\
$[fb^{-1}]$&  $[\TeV]$       & $[\TeV]$      &  $[\TeV]$      & $[\TeV]$      \\
   \hline
   \hline
       1   &   22.1 &   19.0 &    22.1 &   19.0 \\
   \hline                                       
      10   &   31.8 &   24.3 &    31.7 &   24.2 \\
   \hline                                       
     100   &   56.9 &   32.0 &    51.7 &   31.0 \\
   \hline                                       
  \end{tabular}
 \end{center}
\end{table}

The effects of contact interactions are investigated for the $\LL$~model, which
is incorporated in {\tt PYTHIA}. The statistical procedure is as outlined above.
In contrast to the linear collider case, where we have developed a semi-analytical
program with numeric integrations, here we are using a Monte-Carlo generator.
So the task of computing the new physics predictions for a large enough set
of values of the scale $\Lambda$ is much more CPU intensive. We generate 100000
events for each scale value to keep the Monte-Carlo error small and repeat the 
generation 200 times to cover a wide enough range of scales. For the linear
collider studies we use a set with 800 values of the scale for each case, and
they are much faster to compute with a semi-analytical program.

\begin{figure}[!ht]
  \begin{center}
    \resizebox{0.85\textwidth}{12.0cm}{\includegraphics{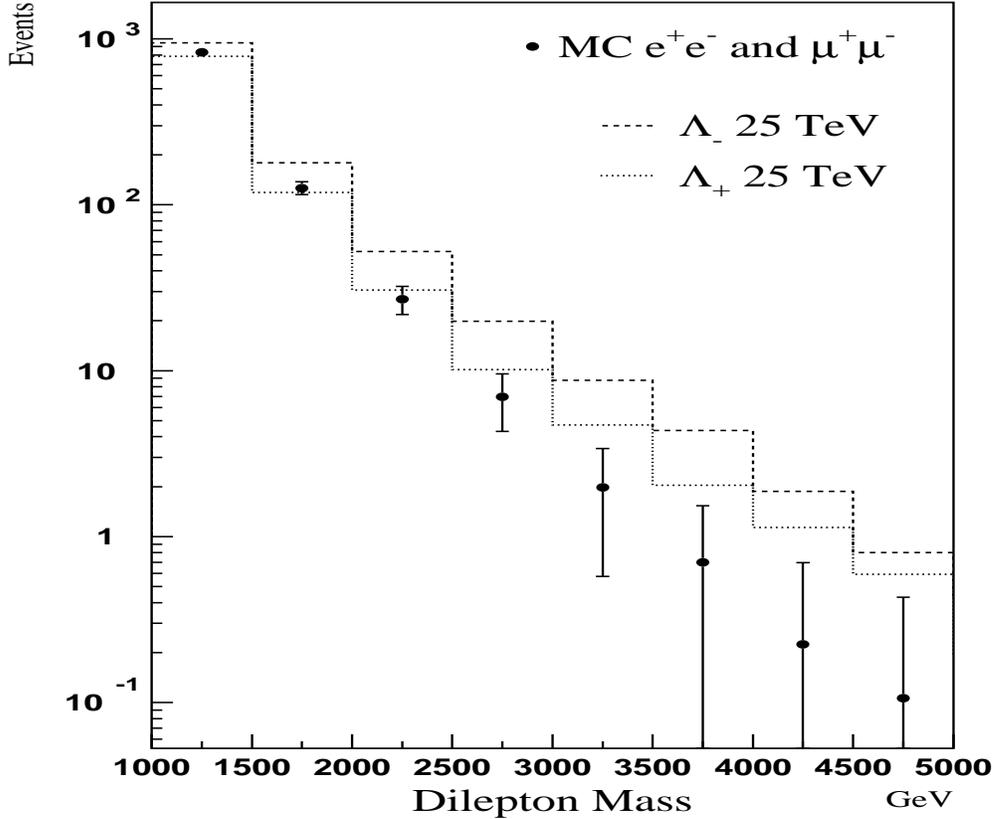}}
  \end{center}
  \caption{\em Effects of contact interactions on the dilepton mass spectrum.
             Points are simulated events in the SM, and the histograms show the spectrum
             in the presence of LL contact interactions with different signs of
             the amplitudes.}
  \label{fig:dy-ci-ll}
\end{figure}

The sensitivity reaches from our fits for different luminosities are summarized
in~\Tabref{dyci-limits}, assuming that we detect electron- and muon-pairs in
one LHC experiment. The sensitivity is dominated by the cross section
measurement, the contribution of the forward-backward asymmetry is minor
due to the large statistical errors and the need to apply rapidity cuts.
Clearly ATLAS and CMS can combine their data.
The sensitivity ranges from 19.0 to 32.0~$\TeV$ for positive and
from 22.1 to 56.9~$\TeV$ for negative interference. 
The effects of contact interactions on the Drell-Yan mass spectrum
are illustrated in Figure~\ref{fig:dy-ci-ll}.
Compared to the sensitivity
reach for hadronic final states at a linear collider~\cite{Sabine:2001},
the positive interference case is similar and the negative interference case
more sensitive. But a linear collider can detect also specific flavors,
e.g. beauty or charm final states, where the sensitivity is higher. So an
indication of something new from LHC can be studied in much greater detail
at a FLC. Even at the highest luminosities the statistical errors 
at LHC are dominant, as evident from the comparison
of the cases with total systematic uncertainties of 3 and 6~\%. 
This is not surprising as the Drell-Yan process is probing directly masses up 
to $\sim$~3--4~$\TeV$, where due to the steeply falling cross sections the
statistical errors dominate by far.

\subsection{$e^{+}e^{-}$ at a Linear Collider}

The effects of new physics at a linear collider with \mbox{$\rs = 0.5\ \TeV$}
are computed with a semi-analytical
program in the improved Born approximation, using effective couplings. QED effects
in the initial and final states are taken into account.
Events without substantial energy loss due to initial state radiation are
selected by a cut on the ``effective'' energy: $\sqrt{\SP} / \sqrt{s} > 0.85$.
For them the interactions occur close to the nominal machine energy and offer
the best sensitivity for manifestations of new physics.
We repeat the computation for each case under study (e.g. one of the eight contact
interaction models) for a set of 800 values of the relevant scale.
The sensitivity to contact interactions is determined by a fit as outlined above.

\begin{table}[htb]
\renewcommand{\arraystretch}{1.20}
\caption{Sensitivity reach for contact interactions at 95\% CL at a linear collider
with \mbox{$\rs = 0.5\ \TeV$} and 1000 fb$^{-1}$.}
\label{tab:ci-limits}
 \begin{center}
  \begin{tabular}{|c|c|c||c|c|}
   \hline
   \multicolumn{5}{|c|}{\boldmath $e^{+}e^{-} \rightarrow e^{+}e^{-}$\unboldmath} \\
   \multicolumn{5}{|c|}{\boldmath Contact Interactions \unboldmath} \\
 & \multicolumn{2}{c}{\boldmath ``Realistic'' \unboldmath} & \multicolumn{2}{c|}{\boldmath Optimistic \unboldmath} \\
   \hline
   Model  &  $\Lambda^{-}$  & $\Lambda^{+}$ &  $\Lambda^{-}$ & $\Lambda^{+}$ \\
          &  $[\TeV]$       & $[\TeV]$      &  $[\TeV]$      & $[\TeV]$      \\
   \hline
   \hline
   LL    &   23.2 &   23.3 &    43.5 &   44.9 \\
   \hline                                       
   RR    &   22.5 &   22.5 &    42.1 &   43.4 \\
   \hline                                       
   VV    &   43.9 &   45.2 &    83.3 &   89.1 \\
   \hline                                       
   AA    &   32.5 &   35.0 &    71.9 &   77.1 \\
   \hline                                       
   LR    &   25.2 &   24.4 &    50.7 &   52.4 \\
   \hline                                       
   RL    &   25.2 &   24.4 &    50.7 &   52.4 \\
   \hline                                       
   LL+RR &   32.0 &   32.6 &    59.9 &   63.0 \\
   \hline                                       
   LR+RL &   35.0 &   35.2 &    71.0 &   75.0 \\
   \hline
  \end{tabular}
 \end{center}
\end{table}

Two cases are distinguished:
\begin{enumerate}
 \item ``Realistic'': a cross section error is composed of the statistical error and
a systematic error of 0.5 \% coming from the experiment, 0.2 \% from the luminosity
determination and a theoretical uncertainty of 0.5 \%. The forward-backward
asymmetry error consists of the statistical error and a systematic uncertainty of
0.002 (absolute) for $e^{+}e^{-}$ and 0.001 (absolute) for $\mu^{+}\mu^{-}$ final
states. The main origins of the latter are from charge confusion of the leptons
and uncertainties in the acceptance edge determination. Both of these effects are
more important for electrons due to the forward peak in the differential cross
section and the longer lever arm for measuring the muon momenta. One should
stress that the forward-backward asymmetry systematics is lower than what was
achieved at LEP, and requires a substantially improved detector.
 \item Optimistic: a cross section error is composed just of the statistical error and
a 0.2 \% contribution from the luminosity determination.
The forward-backward asymmetry error consists of the statistical error and a 
systematic uncertainty which is given by the {\it minimum} of the systematic
uncertainty for the ``Realistic'' case and the statistical error. The rationale 
behind is the hope that with increasing statistics one can control the systematic
effects better. In practice this turns out to play a role only for $e^{+}e^{-}$.
For $\mu^{+}\mu^{-}$ the statistical error of the forward-backward asymmetry is
always bigger than 0.001.
\end{enumerate}

The optimization of the acceptance range is an important experimental question.
The strong forward peak of Bhabha scattering is less sensitive to new physics
as the SM amplitudes dominate the interference terms. We have investigated
two regions:
\begin{itemize}
 \item barrel: from 44$^{\circ}$ to 136$^{\circ}$, where the polar angle is with
       respect to the electron beam line
 \item barrel + backward endcap: from 44$^{\circ}$ to 170$^{\circ}$, so the
       region of backward scattering is added.
\end{itemize}
For contact interactions there is no gain from going outside of the barrel
region, even the sensitivity reach is reduced by 1--2 \%. As the detector
performance is usually highest, and the backgrounds lowest in the central
region, this is the optimal measurement area from experimental point of view.

\begin{table}[htb]
\renewcommand{\arraystretch}{1.20}
\caption{Sensitivity reach for contact interactions at 95\% CL at a linear collider
with \mbox{$\rs = 0.5\ \TeV$}
for the VV model with positive interference as a function of the accumulated luminosity.}
\label{tab:ci-lumi}
 \begin{center}
  \begin{tabular}{|r|c||c|}
   \hline
   \multicolumn{3}{|c|}{\boldmath $e^{+}e^{-} \rightarrow e^{+}e^{-}$ \unboldmath} \\
   \multicolumn{3}{|c|}{\boldmath VV Model \unboldmath} \\
 & \multicolumn{1}{c}{\boldmath ``Realistic'' \unboldmath} & \multicolumn{1}{c|}{\boldmath Optimistic \unboldmath} \\
   \hline
Luminosity  & $\Lambda^{+}$&  $\Lambda^{+}$ \\
$[fb^{-1}]$ & $[\TeV]$     &  $[\TeV]$      \\
   \hline
   \hline
    1       &     27.3     &      28.4      \\
   \hline                                   
   10       &     39.8     &      49.9      \\
   \hline                                   
  100       &     44.4     &      74.8      \\
   \hline                                   
 1000       &     45.2     &      89.1      \\
   \hline                                       
  \end{tabular}
 \end{center}
\end{table}

The sensitivity reaches from our fits for the different models are summarized
in~\Tabref{ci-limits}. They range from 22.5 to 45.2~$\TeV$ in the 
``Realistic'' scenario and are factor
of two higher in the Optimistic case, which should be taken as
an ``ideal'' upper sensitivity limit. The estimates for muon
pairs in~\cite{Sabine:2001} are somewhat higher for 1000~fb$^{-1}$. They are
derived under more optimistic assumptions and use in addition the
left-right asymmetry, which is not included in the present study.
The limits from Bhabha scattering at LEP are factor of 2.6--2.9
lower~\cite{Bourilkov:2000ap,Bourilkov:2001pe,Abbaneo:2002}. 

The sensitivity reach for e.g. the VV model evolves from 27.3~$\TeV$ at start-up to
44.4~$\TeV$ for 100~fb$^{-1}$, where it is saturated by the systematic
effects as depicted in~\Tabref{ci-lumi}.


\section{Extra Spatial Dimensions}

The development of string theory points to the existence of up
to seven additional dimensions, which are compactified at
very small distances, initially estimated to be \mbox{$\sim 10^{-32}$~m,}
and hence far below the scales probed at high energy colliders.
In a radical proposal~\cite{ADD,ADD2},
the hierarchy problem is dealt with by bringing
close the electroweak scale $m_{EW} \sim 1\; \TeV$ and the
Planck scale $M_{Pl} \sim \frac{1}{\sqrt{G_N}} \sim 10^{15}\; \TeV$.
In this framework the effective four-dimensional $M_{Pl}$ is
connected to a new $M_{Pl(4+n)}$ scale in a (4+n) dimensional
theory:
\Be
 M_{Pl}^2 \sim M_{Pl(4+n)}^{2+n} R^n
\Ee
where there are n extra compact spatial dimensions of radius
$\sim R$, which could be probed at present and future colliders.
This can explain the observed weakness of gravity at large distances.
At the same time,
quantum gravity becomes strong at a scale M of the order of few~\TeV{}
and could have observable signatures.
The attractiveness of this proposal is enhanced by the plethora of expected
phenomenological consequences described  by just a few parameters.

In the production of fermion- or boson-pairs in $e^+e^-$ or $pp$ collisions 
this class of
models can be manifested through virtual effects due to the exchange of
gravitons (Kaluza-Klein excitations).
As discussed in~\cite{Hewett,Giudice,Lykken,Shrock,Rizzo}, the exchange of 
spin-2
gravitons modifies in a unique way the differential cross sections for fermion
pairs, providing clear signatures. These models introduce an effective scale
(cut-off), denoted as $M_s$ in~\cite{Hewett,Rizzo}, as $\L_T$ in~\cite{Giudice}
and again as $M_s$ in~\cite{Lykken}.
The first two scales are connected by the relation $M_s=(2/\pi)^{1/4}\Lambda_T$,
which gives numerically \mbox{$\L_T = 1.1195\;M_s$}. They do not depend on the
number of extra dimensions. In the third case the scale exhibits such a
dependence; the relation to the other scales is given by 
\mbox{$M_s^{HLZ}|_{n=4} = \L_T$} for four extra dimensions.

\begin{table}[htb]
\renewcommand{\arraystretch}{1.20}
\caption{Sensitivity reach for extra dimensions at 95\% CL at a linear collider
with \mbox{$\rs = 0.5\ \TeV$} for
the Hewett scale and positive interference as a function of the accumulated luminosity.}
\label{tab:ll-ed}
 \begin{center}
  \begin{tabular}{|r|c||c|}
   \hline
   \multicolumn{3}{|c|}{\boldmath $e^{+}e^{-} \rightarrow e^{+}e^{-}$ \unboldmath} \\
   \multicolumn{3}{|c|}{\boldmath Hewett Scale \unboldmath} \\
 & \multicolumn{1}{c}{\boldmath ``Realistic'' \unboldmath} & \multicolumn{1}{c|}{\boldmath Optimistic \unboldmath} \\
   \hline
Luminosity  & $M_s$        &  $M_s$         \\
$[fb^{-1}]$ & $[\TeV]$     &  $[\TeV]$      \\
   \hline
   \hline
    1       &     2.6      &      2.6       \\
   \hline
   10       &     3.1      &      3.5       \\
   \hline
  100       &     3.3      &      4.2       \\
   \hline
 1000       &     3.3      &      4.6       \\
   \hline
   \hline
   \multicolumn{3}{|c|}{\boldmath $e^{+}e^{-} \rightarrow \mu^{+}\mu^{-}$ \unboldmath} \\
   \multicolumn{3}{|c|}{\boldmath Hewett Scale \unboldmath} \\
 & \multicolumn{1}{c}{\boldmath ``Realistic'' \unboldmath} & \multicolumn{1}{c|}{\boldmath Optimistic \unboldmath} \\
   \hline
Luminosity  & $M_s$        &  $M_s$         \\
$[fb^{-1}]$ & $[\TeV]$     &  $[\TeV]$      \\
   \hline
   \hline
    1       &     1.6      &      1.6       \\
   \hline
   10       &     2.1      &      2.1       \\
   \hline
  100       &     2.8      &      2.8       \\
   \hline
 1000       &     3.5      &      3.5       \\
   \hline
  \end{tabular}
 \end{center}
\end{table}

We will use the scale $M_s$ of~\cite{Hewett} throughout this paper.
The cut-off scale is supposed to be of the order of the
fundamental gravity scale M in 4+n dimensions.
The results are model-dependent, which is usually expressed by the
introduction of an additional parameter
\Be
\varepsilon = \frac{\lambda}{M_s^4}.
\label{epsed}
\Ee
The value of $\lambda$ is not known exactly, the usual assumption is
$\lambda = \pm 1$ to allow for both constructive and destructive interference
effects.

\begin{figure}[!ht]
  \begin{center}
    \resizebox{0.80\textwidth}{11.0cm}{\includegraphics{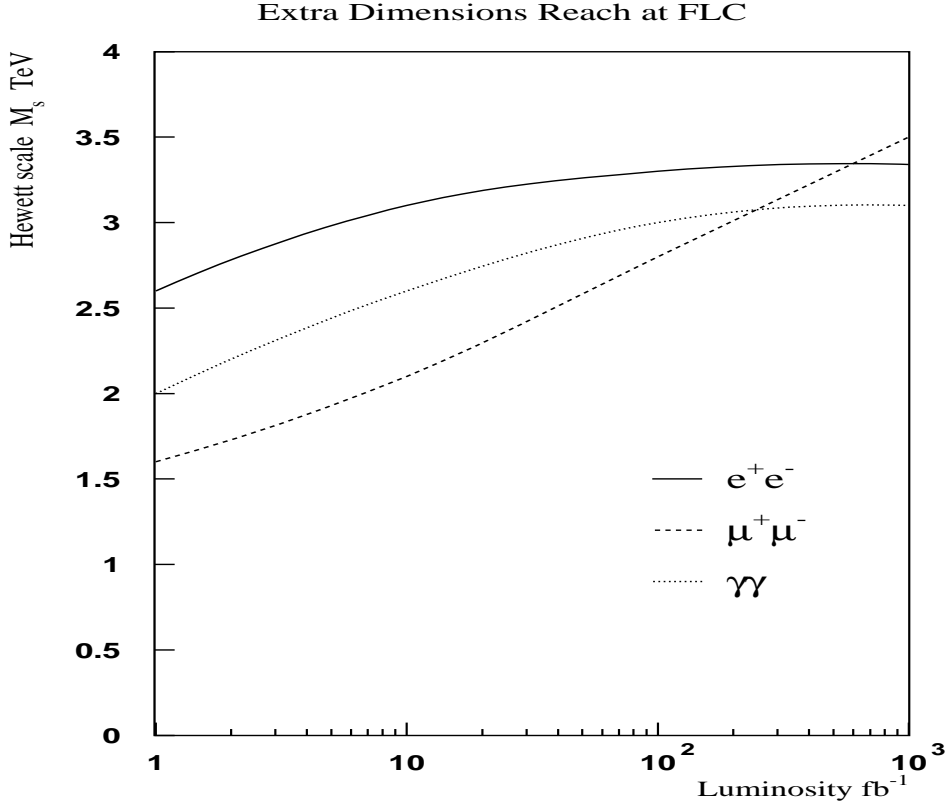}}
  \end{center}
  \caption{\em Evolution of the sensitivity reach for extra dimensions at a linear
           collider with \mbox{$\rs = 0.5\ \TeV$} 
           in different final states with the accumulated luminosity.}
  \label{fig:ed-lumi}
\end{figure}

\subsection{$e^{+}e^{-}$ and $\mu^{+}\mu^{-}$ at a Linear Collider}

The approach is the same as for contact interactions in Bhabha scattering,
as discussed earlier. The parameter $\varepsilon$ is defined in~\Eqref{epsed}.
The virtual graviton effects are computed using the calculations
from~\cite{Hewett,Giudice,Rizzo}.

The sensitivity reaches from our fits for electron- and muon-pairs are summarized
in~\Tabref{ll-ed}. They evolve from 2.6 (1.6)~$\TeV$ for electrons (muons) at
start-up to 3.3 (2.8)~$\TeV$ for 100~fb$^{-1}$. Here the electron channel is
saturated by the systematic uncertainties, while the muon channel is taking over
for the highest luminosities. 
In the tables only the numbers for the positive interference case are shown, as
the sensitivity reach for negative interference is practically the same.
The results are shown in~\Figref{ed-lumi} and agree
with estimates from~\cite{Hewett,Rizzo,Sabine:2001}.

It is interesting to note the large difference between the 
``Realistic'' and the Optimistic scenarios for the two channels: while the electrons
start to saturate above 10~fb$^{-1}$, the muons do not show any saturation at all.
This is explained by the fact that for Bhabhas the sensitivity comes mainly from the
cross section measurement, while for muons is it dominated completely by the 
forward-backward asymmetry, and with our assumptions the statistical error is always
bigger than the systematic uncertainties.

We investigate the optimal acceptance region also for extra dimensions. For
electrons in the final state the gain in sensitivity is below 1~\% from including
the backward endcap, so we may as well restrict the measurement to the barrel
region. For muons the sensitivity comes from the asymmetry which is 
best measured at lower angles, so the results in the table are derived under the
assumption that both the barrel and the two endcaps are used i.e. 
from 10$^{\circ}$ to 170$^{\circ}$.

The present limits on the Hewett scale are $\sim$~1.3~$\TeV$ from LEP and TEVATRON
measurements (see
e.g.~\cite{Bourilkov:2001tp,Landsberg:2001ma,Hewett:2002hv,Cheung:2003ah,Abbaneo:2002}).

\subsection{$\gamma\gamma$ at a Linear Collider}

The production of photon pairs in $\ee$ collisions is described by
$t$- and $u$-channel QED diagrams.
The differential cross section has the following simple form
\vspace{-0.2cm}
\Be
\frac{d \sigma}{d \Omega} = |e_t+e_u +  New\;Physics\;?!|^2
\Ee
\Be
\left(\frac{d \sigma}{d \Omega}\right)_{QED} = \frac{\alpha^2}{2s}\left[\frac{t}{u}+\frac{u}{t}\right] = \frac{\alpha^2}{s}\cdot\frac{1 + \cos^2\theta}{1 - \cos^2\theta}.
\Ee

\begin{table}[htb]
\renewcommand{\arraystretch}{1.20}
\caption{Sensitivity reach for extra dimensions at 95\% CL at a linear collider
with \mbox{$\rs = 0.5\ \TeV$} for the Hewett scale, and for the QED cut-off, in the case of
positive interference as a function of the accumulated luminosity.}
\label{tab:gg-ed}
 \begin{center}
  \begin{tabular}{|r|c||c|}
   \hline
   \multicolumn{3}{|c|}{\boldmath $e^{+}e^{-} \rightarrow \gamma\gamma$ \unboldmath} \\
   \multicolumn{3}{|c|}{\boldmath Hewett Scale \unboldmath} \\
 & \multicolumn{1}{c}{\boldmath ``Realistic'' \unboldmath} & \multicolumn{1}{c|}{\boldmath Optimistic \unboldmath} \\
   \hline
Luminosity  & $M_s$        &  $M_s$         \\
$[fb^{-1}]$ & $[\TeV]$     &  $[\TeV]$      \\
   \hline
   \hline
    1       &     2.0      &      2.0       \\
   \hline
   10       &     2.6      &      2.6       \\
   \hline
  100       &     3.0      &      3.4       \\
   \hline
 1000       &     3.1      &      4.1       \\
   \hline                                       
   \hline                                       
$\L^{QED}$ 1000&  1.2      &      1.6       \\
   \hline                                       
  \end{tabular}
 \end{center}
\end{table}

Deviations from QED typically have the form:
\vspace{-0.2cm}
\begin{eqnarray}
\label{eq11}
\frac{d \sigma}{d \Omega} & = & \left(\frac{d \sigma}{d \Omega}\right)_{QED}\cdot \left(1 \pm \frac{1}{(\L_{\pm}^{QED})^4}\cdot \frac{s^2}{2}\sin^2\theta \right)  \\
\label{eq12}
\frac{d \sigma}{d \Omega} & = & \left(\frac{d \sigma}{d \Omega}\right)_{QED}\cdot \left(1 \pm \frac{\lambda}{\pi\alpha (M_s)^4}\cdot \frac{s^2}{2}\sin^2\theta + ...\right)
\end{eqnarray}
The QED cut-off -~\Eqref{eq11}, is the basic form of possible deviations from
quantum electrodynamics.
 \Eqref{eq12} is the low scale gravity
case~\cite{Giudice,Agashe}.
If we ignore higher order terms (given by $...$),
the equations predict the same form of deviations in the
differential cross section.
In this notation, it is particularly easy to compare the results from
different searches by transforming the relevant parameters.
The relation between the Hewett scale and the QED cut-off is:
$$M_s = 2.57\;\L^{QED}.$$

The sensitivity reach from our fits is summarized in~\Tabref{gg-ed}. The sensitivity
evolves from 2~$\TeV$ at start-up to 3~$\TeV$ for 100~fb$^{-1}$, where
it is saturated by the systematic effects. The LEP limits are 
$\sim$~1 $\TeV$~\cite{SMele:2000,Bourilkov:2001tp,Abbaneo:2002}.
As for Bhabhas, there is no gain in sensitivity going outside of the barrel
region. Here the differential cross section is symmetric, exhibiting both
a forward and a backward peak.

\begin{table}[htb]
\renewcommand{\arraystretch}{1.20}
\caption{Sensitivity reach for extra dimensions at 95\% CL at a linear collider
with \mbox{$\rs = 0.5\ \TeV$} for
the Hewett scale and positive interference as a function of the accumulated luminosity.}
\label{tab:all-ed}
 \begin{center}
  \begin{tabular}{|r|c||c|}
   \hline
   \multicolumn{3}{|c|}{\boldmath $e^{+}e^{-} \rightarrow e^{+}e^{-},\ \mu^{+}\mu^{-},\ \gamma\gamma$ \unboldmath} \\
   \multicolumn{3}{|c|}{\boldmath Hewett Scale \unboldmath} \\
 & \multicolumn{1}{c}{\boldmath ``Realistic'' \unboldmath} & \multicolumn{1}{c|}{\boldmath Optimistic \unboldmath} \\
   \hline
Luminosity  & $M_s$        &  $M_s$         \\
$[fb^{-1}]$ & $[\TeV]$     &  $[\TeV]$      \\
   \hline
   \hline
    1       &     2.6      &      2.6       \\
   \hline
   10       &     3.2      &      3.5       \\
   \hline
  100       &     3.5      &      4.3       \\
   \hline
 1000       &     3.8      &      4.8       \\
   \hline                                       
  \end{tabular}
 \end{center}
\end{table}

If we combine the results from $\gamma\gamma$, $e^{+}e^{-}$ and $\mu^{+}\mu^{-}$,
we get the sensitivity shown in~\Tabref{all-ed}. Scales up to
3.8~$\TeV$ can be probed.


\subsection*{Acknowledgments}

This work is supported in part by the United States National Science Foundation
under grants NSF ITR-0086044 and NSF PHY-0122557.



\end{document}